# Within the MDT Room: Situated in Multidisciplinary Team-Grounded Agent Debate for Clinical Diagnosis

Peng Kuai
Sichuan University
Sichuan, China
2023141460240@stu.scu.edu.cn

Yukun Yang
Tongji University
Shanghai, China
yangyyk@tongji.edu.cn

Shaolun Ruan
Singapore Management University
Singapore, Singapore
slruan.2021@phdcs.smu.edu.sg

Junchi Xu
Sichuan University
Sichuan, China
2024223040091@stu.scu.edu.cn

Yanjie Zhang
The Hong Kong University of Science
and Technology
Hong Kong, China
yzhangvj@connect.ust.hk

Lin Zhang
Zhongnan Hospital of Wuhan
University
Hubei, China

Min Zhu
Sichuan University
Sichuan, China
zhumin@scu.edu.cn

Rui Sheng[*]
The Hong Kong University of Science
and Technology
Hong Kong, China
rshengac@connect.ust.hk

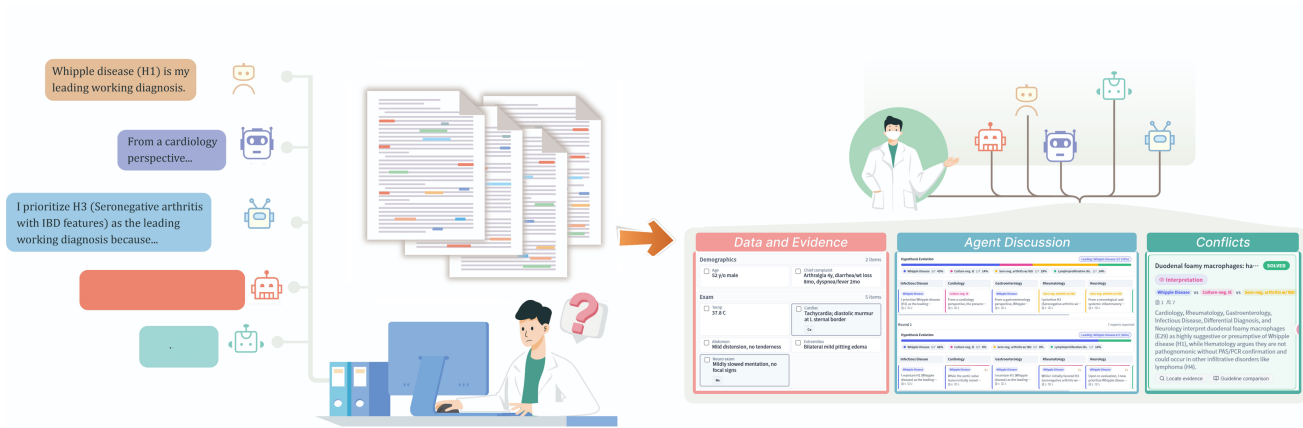

Figure 1: Prior multi-agent debating systems for rare diseases typically present final AI recommendations with long and linear agent discussion logs, offering limited support for clinician inspection of reasoning or timely intervention. *MDTRoom* breaks this paradigm by transforming multi-agent diagnostic debate into a structured, inspectable workspace, which consists of three coordinated views, that enables timely clinician oversight and targeted intervention.

## Abstract

Rare disease diagnosis is inherently challenging due to heterogeneous symptoms, limited clinical familiarity, and fragmented evidence across specialties. Recent large language model (LLM)-based agentic systems have shown promise by simulating multidisciplinary team discussions to generate and evaluate diagnostic hypotheses. However, fully automated diagnosis remains unrealistic, and existing human-in-the-loop approaches provide limited support for effective clinician–agent collaboration. In practice, clinicians are often presented with final diagnostic outputs and lengthy, unstructured agent discussion logs, making it difficult to inspect reasoning, intervene in a timely manner, or guide agent deliberation effectively. To address these challenges, we developed *MDTRoom*, an interactive system that transforms multi-agent discussions from linear transcripts into a structured, inspectable workspace. The system externalizes patient data, evidence provenance, hypothesis evolution, and inter-agent conflicts as interconnected visual objects, enabling clinicians to efficiently examine, intervene in, and guide agent reasoning. Our evaluation demonstrates the effectiveness of *MDTRoom* in supporting clinician–agent collaboration.

**ACM Reference Format:**
Peng Kuai, Yukun Yang, Shaolun Ruan, Junchi Xu, Yanjie Zhang, Lin Zhang, Min Zhu, and Rui Sheng. 2026. Within the MDT Room: Situated in Multidisciplinary Team–Grounded Agent Debate for Clinical Diagnosis. In . ACM, New York, NY, USA, 12 pages. https://doi.org/10.1145/nnnnnnn.nnnnnnn

## 1 Introduction

Rare diseases, collectively affecting over 300 million people, often present with heterogeneous, multisystem symptoms that cross specialty boundaries, making diagnosis challenging [15, 29]. For example, conditions like Whipple disease, which mimic more common disorders, frequently lead to years-long diagnostic delays [14, 43]. Recently, large language model (LLM)-based agentic systems have been proposed to support rare disease diagnosis. One of the most common approaches is to introduce multiple LLM-based agents to simulate multidisciplinary team (MDT) consultations, where each agent represents a specialist and engages in structured discussion and debate to generate differential diagnoses [6, 8, 45, 54]. This approach has been implemented in systems such as RareAgents [6] and MAC [8], which demonstrate strong diagnostic performance and increasingly traceable reasoning.

Despite these advances, fully automated rare disease diagnosis remains unrealistic, as agentic systems are prone to hallucinations [18] and raise ethical concerns. Consequently, clinicians often treat these systems as decision-support tools rather than autonomous decision- makers [44, 51]. In such challenging rare disease diagnosis scenarios, clinicians can rely on the system's ability to generate diagnostic hypotheses and evaluate them based on literature evidence. This can help surface potential diagnoses that might otherwise be overlooked. At the same time, clinicians can leverage their knowledge to guide debates among multiple agents and promote more efficient collaboration with those agents.

However, existing human-in-the-loop approaches for multi-agent debate provide little support for fostering such clinician–agent collaboration. Specifically, current systems mainly provide users with the final diagnostic results and the extensive log data generated during agent discussions [9, 10, 36]. Clinicians would review these discussions to adjust their judgments based on agents' reasoning in the context of rare disease diagnosis. However, the sheer volume of the data and the back-and-forth reasoning across multiple agent roles make this process cognitively demanding. Consequently, clinicians may miss critical information, such as areas of high diagnostic uncertainty or emerging alternative hypotheses. In addition, clinicians may be unable to inject their domain knowledge in a timely manner, causing them to spend effort reviewing agent reasoning that has already drifted away from clinically relevant directions. As a result, limited clinician–agent collaboration can lead to delayed or suboptimal diagnostic decisions.

To address this problem, we first conducted a formative study with eight experienced clinicians and derived six design requirements to help clinicians more efficiently examine, intervene in, and guide agent debates. Then we proposed *MDTRoom*, an interactive system to support clinician-agent collaboration in rare disease diagnosis. The central design idea of our system is to transform multi-agent discussion from linear transcripts into a structured, inspectable workspace. Specifically, in this workspace, original patient data, evidence provenance, hypothesis evolution, and inter-agent conflicts are externalized as interconnected visual objects. This allows clinicians to effectively navigate, understand, and audit the reasoning process without having to piece it together from text. The main contributions of this paper are as follows.

- We conduct a formative study with eight experienced clinicians to identify the practices and challenges of clinician–multi-agent collaboration in rare disease diagnosis.
- We propose an interactive tool, *MDTRoom*, designed to support clinicians in examining, auditing, and guiding multi-agent reasoning processes.
- We evaluate our system through a user study, demonstrating that it enhances clinicians' understanding of multi-agent reasoning, increases engagement, and reduces cognitive load.

## 2 Related Works

In this section, we review three closely related lines of work that together contextualize the design and contributions of *MDTRoom*. Specifically, we first survey recent advances in LLM-driven approaches for rare disease diagnosis, which motivate the use of large language models and multi-agent reasoning in complex, long-tail clinical scenarios. We then review research on human–AI collaboration for clinical decision-making. Finally, we examine interactive systems for inspecting and steering multi-agent behaviors, which inform our interface design but have largely been developed outside of clinical settings. By synthesizing insights from these three strands, we position *MDTRoom* at their intersection and clarify how it addresses limitations not fully explored in prior work.

### 2.1 LLM-driven Rare Disease Diagnosis

LLM-driven diagnosis has gained increasing attention in recent years [26, 46, 48]. A growing body of work has explored LLM-driven approaches for rare disease diagnosis [6–8, 37, 54]. First, some works focus on adapting and enhancing large foundation models as standalone diagnostic assistants [7, 28, 37, 38]. Concretely, frontier medical LLMs have achieved competitive results across major clinical benchmarks and standardized differential-diagnosis tasks [28, 38]. This motivates further investigation into whether these capabilities extend to the long-tail and data-sparse setting of rare diseases. Some research also considers improving LLMs through structural augmentation and the incorporation of external medical knowledge [2, 7]. Rather than relying solely on the foundation model's internal representations, these approaches explicitly scaffold the reasoning process or enrich the input with domain-specific signals. More recent work moves beyond single-model improvement and formulates diagnosis as a multi-agent process. In the general medical domain, systems such as AI Hospital [12], MedAgents [41], and MDAgents [20] simulate diagnosis using multiple agents with different roles, flexible collaboration patterns, or hospital-like interactive settings, allowing models to reason in a way that more closely resembles real clinical practice. Specifically, in rare disease scenarios, this idea has been further developed into MDT-inspired multi-agent systems, including MAC [8], RareAgents [6], and DeepRare [54]. These systems rely on multiple specialist agents, tool use, and step-by-step discussion to reduce missed diagnoses and improve performance on complex and uncommon cases. Despite these advances, prior work largely emphasizes algorithm improvements, while leaving the interaction layer underexplored. In particular, little is known about how clinicians can inspect multi-agent deliberations, interpret the use of evidence and guidelines, or intervene when reasoning fails. Our work focuses on addressing this interaction gap.

## 2.2 Human–AI Collaboration for Diagnosis

Given the high stakes and complexity of clinical diagnosis, a growing body of work has focused on designing effective clinician–AI collaboration, rather than treating AI systems as fully autonomous decision-makers [3, 5, 23, 35]. Accordingly, those systems have been developed to support different stages of clinical diagnosis, such as trajectory exploration, augmented report review, visual querying, or conversational assistance [16, 19, 25, 33, 40, 42, 52]. In AI systems that can directly generate clinical decisions, much of the research has emphasized explainability methods to support transparency and trust in high-stakes clinical settings [1, 22, 30]. Two types of explainability approaches are commonly studied [35]. The first is feature attribution, which indicates the supporting or opposing role of individual input features in shaping the model's prediction. In healthcare, feature attributions can reveal how specific lab values, clinical signs, and other recorded observations influence a given prediction, helping clinicians judge whether the model attended to clinically relevant findings [21, 30, 39]. For example, Krause et al. [21] developed an interactive system that allowed clinicians to inspect the most important features for a given patient's prediction, while Sivaraman et al. [39] displayed SHAP-based feature attribution charts to help ICU clinicians understand how patient states were characterized by the model. In addition, case-based explanations present similar or contrastive historical cases alongside the current prediction, allowing clinicians to assess plausibility by analogy [5, 50]. For instance, CheXplain [50] displayed counterfactual chest X-ray images, as physicians reported that contrasting normal with abnormal images was a routine part of their diagnostic workflow. With the advent of LLMs, natural language itself has become a medium for agent reasoning: models can directly genaerate step-by-step diagnostic reasoning in prose, articulating which findings support or argue against each hypothesis [28, 38]. Recent studies have also shown that clinicians respond differently depending on whether AI support is delivered as blunt predictions, actionable intermediate guidance, or LLM-augmented responses with citations and contextual explanations [34, 53]. However, in multi-agent diagnostic debates, the key challenge is no longer explaining a single model's output, but enabling clinicians to inspect a distributed deliberation process involving multiple agents, their evidence, and points of disagreement.

## 2.3 Human–Multi-agent Interaction Systems

Adjacent work has begun to explore how users inspect and steer LLM-based multi-agent systems, primarily from developer-facing and end-user perspectives. For example, AutoGen Studio [10] provides a no-code interface for building and monitoring multi-agent workflows, streaming agent messages in real time. AGDebugger [11] targets post-hoc debugging of multi-agent failures, enabling developers to replay execution traces and test alternative agent configurations. DiLLS [36] introduces a layered summary framework grounded in Activity Theory that organizes agent behaviors into three levels, significantly improving developers' ability to locate and understand failures in centralized multi-agent systems. Beyond debugging, AgentLens [27] offers a visual analytics approach for exploring agent behaviors in LLM-based autonomous simulations, using hierarchical temporal visualization and causal tracing to help end users understand how agent events evolve over time. AgentCoord [31] takes a complementary angle by supporting the visual design and exploration of coordination strategies for multi-agent collaboration, allowing users to interactively edit agent roles, task dependencies, and execution flows. These systems collectively demonstrate that structured visual interfaces are essential for making multi-agent behaviors comprehensible [32, 49]. However, they are not tailored to clinical use. In rare disease diagnosis, clinicians must interpret agent discussions under high diagnostic uncertainty and significant clinical responsibility. Existing systems do not adequately support these needs, leaving clinicians to manually sift through complex agent interactions without effective mechanisms for timely examination and guidance.

## 3 Formative Study

To inform the design of *MDTRoom*, we conducted a formative study with eight experienced clinicians. Our goal was to understand the challenges clinicians face when collaborating with an LLM-based multi-agent debate system in the context of rare disease diagnosis, as well as the forms of support needed to help them inspect, guide, and correct the underlying reasoning process.

### 3.1 Participants and Procedure

We interviewed eight clinicians (E1–E8; 6 female, 2 male) recruited through advertisements posted on social media platforms. The participants spanned diverse specialties: neurology (E1), cardiology (E2), critical care (E3), veterinary medicine (E4), laboratory medicine (E5), oral medicine (E6), pulmonology (E7), and dermatology (E8). All participants had between 5 and 7 years of professional experience in clinical diagnosis and encountered rare disease cases on a monthly basis. Furthermore, the eight clinicians had prior experience using LLMs for clinical diagnosis, such as biomedical literature retrieval and differential diagnosis hypothesis generation. Detailed demographic information is shown in Table 1. To familiarize participants with the multi-agent debate paradigm and the system, we conducted a brief training session. Participants were introduced to a representative LLM-based multi-agent debating system [24] and the overall workflow, and then asked to explore several representative rare disease cases based on their domain knowledge. Then, we conducted semi-structured interviews to understand the challenges participants encountered during their collaboration with the existing LLM-based multi-agent system. All sessions were audio-recorded with participants' consent, and each participant received around US$12 as compensation. This study was approved by the institutional ethics committee.

**Table 1: Participant demographics in the formative study.**

| ID | Specialty | Gender | Experience | AI Usage |
|----|-----------|--------|------------|----------|
| E1 | Neurology | F | 7 years | Monthly |
| E2 | Cardiology | F | 7 years | Monthly |
| E3 | Critical Care | F | 7 years | Weekly |
| E4 | Veterinary Med | F | 6 years | Monthly |
| E5 | Laboratory Med | F | 5 years | Daily |
| E6 | Oral Medicine | M | 7 years | Daily |
| E7 | Pulmonology | M | 5 years | Daily |
| E8 | Dermatology | F | 7 years | Monthly |

## 3.2 Diagnostic Challenges

We examined the interview data using a thematic analysis approach [4]. Two researchers separately reviewed and annotated the transcripts, with attention to difficulties clinicians encountered during the exploration tasks. They subsequently compared their interpretations, resolving inconsistencies through discussion and by referring back to the original audio recordings. Through iterative synthesis, related observations were consolidated into higher-level themes, from which we distilled three recurrent diagnostic challenges in clinician collaboration with multi-agent debate systems.

**(C1) Difficulty following distributed reasoning across multiple agents.** All eight participants reported difficulty tracking individual agents' conclusions, how those conclusions evolved over time, and which pieces of clinical evidence were used to support them. Agent outputs were presented as lengthy, unstructured text streams that interleaved evidence citations, intermediate inferences, and diagnostic hypotheses without clear separation. This difficulty was further exacerbated by the fact that each agent reasoned over partially overlapping subsets of the clinical data, often emphasizing different findings. As a result, clinicians struggled to determine which evidence had been considered by which agent and how selectively it had been interpreted. For example, E4 mentioned, *"Each agent jumps between findings and conclusions. I cannot tell which evidence actually supports a given diagnosis unless I reconstruct the reasoning myself."* E1 also noted that organizing unstructured dialogues into structured summaries is itself laborious, and wished the system could assist rather than adding to that burden.

**(C2) Difficulty detecting and understanding conflict and consensus among agents.** Seven participants (E1–E5, E7, E8) reported that identifying inter-agent disagreement was clinically valuable but difficult to detect in the prototype. Recognizing disagreement helps clinicians pinpoint areas of diagnostic uncertainty, which is particularly important in rare disease diagnosis, where evidence is sparse and symptoms are atypical. E2 noted that in real MDT meetings, disagreement is immediately apparent because dissenting opinions interrupt the conversational flow; in the prototype, however, conflicting views were silently embedded in adjacent paragraphs without any visual differentiation. At the same time, they also sought visibility into agent agreement, which strengthens diagnostic confidence and supports hypothesis prioritization. Additionally, three participants (E1, E3, and E5) highlighted that knowing which data each agent used and how they interpreted it would help them understand why agents reached different conclusions. For example, E4 thought that different specialists would *"focus on different pieces of information"* or even *"obtain different insights from the same evidence"*. However, due to the large volume and heterogeneity of patient data, it is challenging to track exactly how each specialist uses the data, which in turn makes it difficult to understand the sources of conflict.

**(C3) Difficulty integrating domain knowledge effectively and in a timely manner.** Six participants (E1, E2, E4, E6, E7, E8) expressed a strong need to intervene directly and selectively when agent reasoning deviated from clinical expectations. In the current systems, however, agents generate outputs rapidly and in large volumes, making it difficult for clinicians to follow the reasoning process and provide timely guidance. As E5 noted, *"By the time I notice the agent has misinterpreted a test, it has already produced several downstream suggestions. Therefore, I cannot correct it fast enough."* Clinicians also emphasized the need to supply raw clinical data to agents selectively. E3 explained, *"Sometimes I want to feed a specific lab result to just one agent, not the whole group, so it updates only the relevant reasoning chain."* In addition, participants noted that the current process for providing data is cumbersome, often requiring manual copying and pasting, which is especially burdensome when data are dispersed across multiple sources. However, existing systems provide limited support for such targeted interactions, offering few mechanisms for real-time data injection or intervention. E7 also stated, *"Without being able to intervene precisely and immediately, I end up spending a lot of time checking things that are already going off-track."*

## 3.3 Design Considerations

Based on our findings from the interview, we derived five design considerations to support more effective clinician–multi-agent collaboration in rare disease diagnosis.

**DC1 Structure multi-agent reasoning process.** To address clinicians' difficulty in following distributed reasoning across multiple agents (C1), the system decomposes each agent's outputs into structured components, making it clear which evidence was cited, what inferences were drawn, and which hypotheses were supported. Instead of presenting reasoning as long-form, interleaved text, clinicians can inspect agent-specific chains directly, reducing the cognitive effort required to reconstruct each agent's contribution.

**DC2 Make conflict and consensus explicit.** Recognizing that inter-agent conflict and consensus are highly informative yet often invisible in text-centric outputs (C2), the system should proactively highlight points of divergence and convergence, showing both the evidence and reasoning behind each stance. Clinicians can therefore quickly assess competing interpretations and gauge diagnostic confidence without manually comparing paragraphs across agents.

**DC3 Highlight agent-specific data and evidence usage.** To support clinicians in understanding why agents reach different conclusions (C2), the system should make explicit which raw data each agent attended to and how it was interpreted through the clinical evidence. By visualizing agent-specific data usage, clinicians can more easily track the reasoning process, identify the sources of disagreement, and assess the reliability of each agent's conclusions.
**DC4 Support targeted intervention.** Clinicians face difficulty integrating domain knowledge effectively and in a timely manner (C3), particularly when agents generate outputs rapidly and in large volumes. To address this, our system should allow clinicians to pause or terminate individual agent discussions at any time and selectively guide their reasoning. In addition, to facilitate data-grounded intervention, the system should enable clinicians to provide raw patient data for in-context prompting and direct it to specific agents as needed. This design supports precise, timely guidance, reduces cognitive load, and prevents the propagation of local errors across the multi-agent system.
**DC5 Support comparative traceability after clinician-guided interventions.** After clinicians inject their own domain knowledge to guide model reasoning, the system should explicitly surface how the discussion evolves relative to prior states (C3). Specifically, it should preserve the full temporal history while visually highlighting only the localized changes introduced by clinician interventions, including shifts in structure, stance, and conclusions. This enables clinicians to directly compare pre- and post-intervention reasoning without overwriting earlier discussions, thereby supporting reflective review, retrospective auditing, and informed debate grounded in historical context.

## 4 MDTRoom

Guided by **DC1**–**DC5**, we designed *MDTRoom* as an interactive workspace where clinicians collaborate with a multi-agent diagnostic system in real time. Rather than leaving clinicians to read through lengthy discussion transcripts, *MDTRoom* structures the agent discussion process, from evidence interpretation to conflict resolution, into coordinated visual representations that can be inspected, navigated, and corrected at every stage.

The system begins with a case grounding and agent configuration phase. In the first step, clinicians describe a patient case in free-form natural language, reflecting how clinical information typically arrives in practice, often as narrative notes and referral letters rather than structured records. In the second step, the system uses Qwen3-Max[1] to parse the narrative into editable structured items, such as *Symptoms*, *Labs*, and *Imaging*). Clinicians can revise, add, or remove any extracted item, establishing a shared case representation that all subsequent views reference. In the third step, clinicians can select participating AI agents from a pool spanning organ systems and diagnostic roles, mirroring how referring clinicians in real MDT meetings decide which specialists to invite based on the demands of each case.

Once the case is configured, the interface transitions to a three-panel discussion workspace anchored by a shared round timeline (Figure 3) (**DC1**). The overall workflow proceeds as follows. The *Report and Evidence View* (Figure 3-A) keeps the patient's clinical data visible, enabling clinicians to continuously verify that agent reasoning remains grounded in the original data and inspect how agents leverage specific clinical literature as evidence. The *Agent Discussion View* (Figure 3-B) presents the multi-agent debate organized by round, enabling clinicians to monitor how each AI specialist's opinion evolves and where opinion changes occur. The *Conflict View* (Figure 3-C) extracts disagreements from the discussion and surfaces them as explicit, actionable objects, allowing clinicians to triage which conflicts require their attention.

[1]https://qwen.ai/blog?id=241398b9cd6353de490b0f82806c7848c5d2777d&from=research.latest-advancements-list

### 4.1 Report and Evidence View

The *Report and Evidence View* (Figure 3-A) presents the raw data alongside key clinical evidence that supports agent reasoning during the discussion process. It has two primary purposes: first, to let clinicians verify at any moment whether the agents' reasoning remains grounded in the original patient data and appropriate clinical evidence; second, to enable clinicians to instruct or correct agents directly based on their domain knowledge.

First, this view presents the patient's raw clinical data (e.g., vital signs, laboratory values, imaging reports, and clinical history) in a structured format (e.g., *Demographics*, *Exam*, *History*, *Labs*, *Imaging*). These data items remain persistently displayed throughout the agent discussion, ensuring that clinicians always have access to the original patient information rather than relying on agents' summaries of their data usage. To make agent reasoning traceable, each data item is annotated with small colored badges (Figure 3-a1) indicating which agents have relied on it and for which diagnostic hypothesis (**DC3**). For each referenced data item, the view also displays the clinical evidence (e.g., clinical best practices or biomedical literature) that agents leveraged to justify their opinion, making the reasoning chain explicitly inspectable. When two or more agents rely on the same data item but reach different diagnostic conclusions, the item will be flagged with a *Conflict* badge (Figure 2-left). Once the disagreement is resolved in subsequent rounds, the badge changes to *Resolved* (Figure 2-right), providing a visual record of how conflicts around specific data items were settled over time.

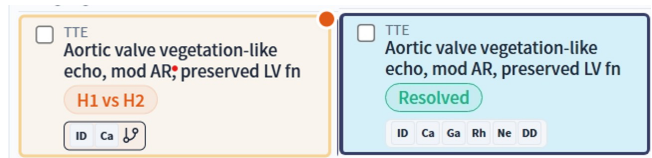

**Figure 2: Data provenance annotations. Left: an imaging finding (data item) with agent badges showing which specialists relied on it and for which diagnosis hypothesis. Right: the same item after the disagreement was resolved, indicated by the *Resolved* badge.**

At the bottom of this view, a clinical intervention module (Figure 3-a3) allows clinicians to inject their expertise into the discussion (**DC4**). Clinicians can select one or more data items from the list above, provide a free-text clinical instruction, and specify which agents should receive the guidance. This will trigger a revision round: the targeted agents will re-evaluate their reasoning in response to the provided instructions, and their updated responses appear in the shared timeline while other agents' contributions are preserved (**DC5**).

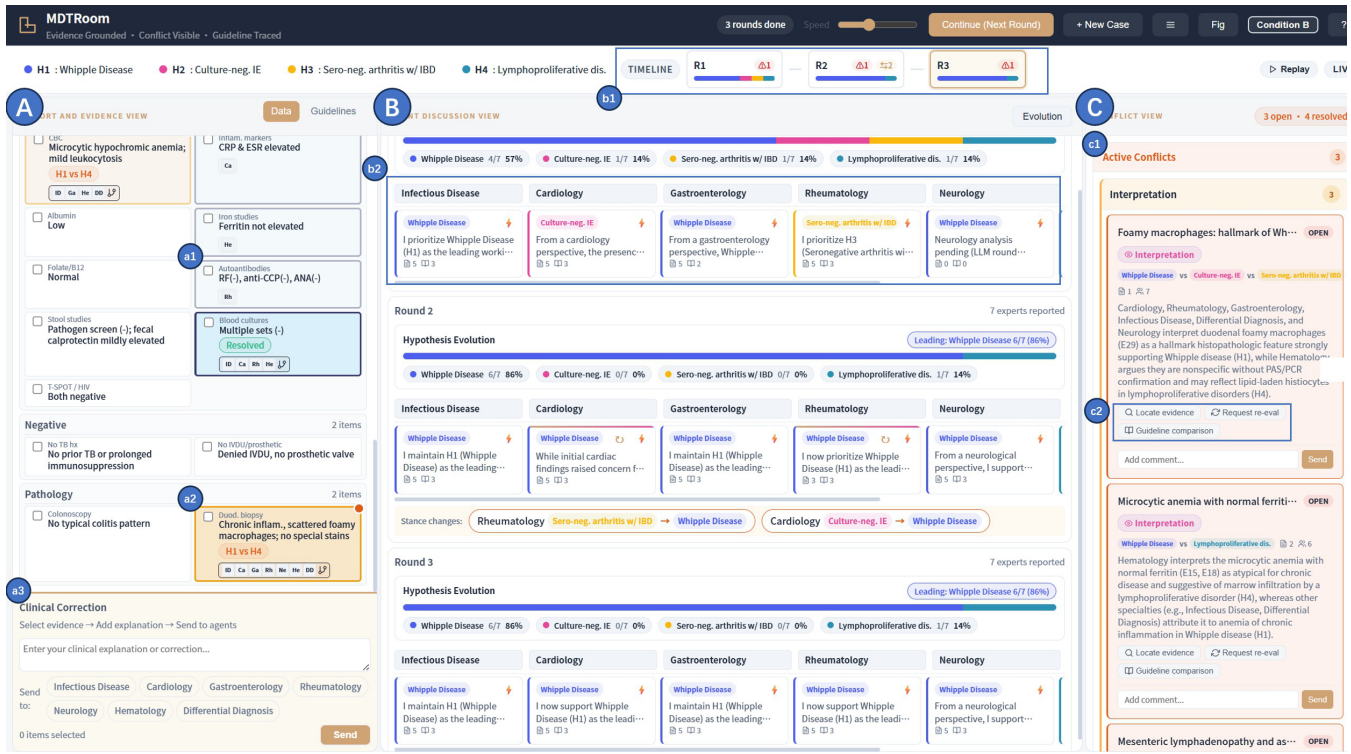

Figure 3: The *MDTRoom* discussion workspace. (A) The *Report and Evidence View* displays patient data with provenance annotations and provides a clinical intervention module at the bottom. (B) The *Agent Discussion View* organizes the multi-agent debate by round, with a shared timeline at the top and agent opinion cards below. (C) The *Conflict View* surfaces disagreements as explicit objects separated into active and resolved groups, with a consensus summary below.

## 4.2 Agent Discussion View

The *Agent Discussion View* (Figure 3-B) is where the multi-agent debate unfolds. Its primary purpose is to help clinicians both monitor the discussion as it progresses and trace how each specialist's opinion has changed over time.

At the top of this view, a shared round timeline (Figure 3-b1) summarizes the overall progress of the discussion. Each round is represented as a card containing a stacked bar that shows how support is distributed across competing hypotheses, together with badges indicating new conflicts or opinion changes in that round. Clinicians can click any round card to revisit that point in the discussion and inspect the state as it was, then return to the current state. The system also supports revision rounds: when a clinician submits an intervention prompt through the *Report and Evidence View*, the targeted LLM-based agent specialists will produce updated responses that appear as a new round in the same timeline, keeping the full discussion history intact.

Furthermore, the debate is organized round by round (Figure 3-b2). Each round displays agent *opinion cards* (one per specialist), showing the agent's current hypothesis, the number of data items and clinical evidence it referenced, and whether it changed its opinion from the previous round. We chose this card-based layout considering that clinicians hope to check individual agent specialists' opinions, for example, what the cardiologist concluded, or whether the infectious disease specialist agreed, rather than reading through a long sequential discussion (**DC1**). Clicking an opinion card opens a detail panel that reveals the agent's stepwise reasoning chain, summary, and cited evidence. Between rounds, an opinion-change strip summarizes which agents switched hypotheses and how their hypotheses changed, allowing clinicians to quickly identify shifts without opening individual cards. This can help clinicians quickly identify opinion shifts. To help clinicians assess whether the group is converging toward a shared diagnosis or remaining divided, the view also offers an optional flow visualization that tracks each agent's opinion across all rounds.

## 4.3 Conflict View

This information is particularly valuable in rare disease diagnosis, where ambiguous evidence often leads to multiple plausible interpretations, and specialist disagreements highlight areas of greatest diagnostic uncertainty requiring clinician judgment (**DC2**). Rather than leaving these disagreements buried in discussion text, this view extracts them and presents them as explicit, actionable objects.

The view separates conflicts into *Active* and *Resolved* groups (Figure 3-c1). This helps clinicians focus on unresolved disagreements and reduces unnecessary re-review. The transition of conflicts from active to resolved further serves as a natural indicator of discussion convergence. Each conflict card displays the agents involved and the competing hypotheses as color-coded tags matching the global hypothesis legend. Selecting a conflict cross-highlights the disputed data items in the *Report and Evidence View* and the relevant agent messages in the *Agent Discussion View*, helping clinicians trace any disagreement across the entire workspace. Each card also provides action buttons (Figure 3-c2): *Locate evidence* highlights the disputed data items, *Request re-eval* sends a targeted re-evaluation request to the involved agents, and *Evidence comparison* shows how the disagreeing agents each applied clinical evidence, helping clinicians pinpoint whether the conflict stems from different evidence or different interpretation of the same evidence (**DC3**).

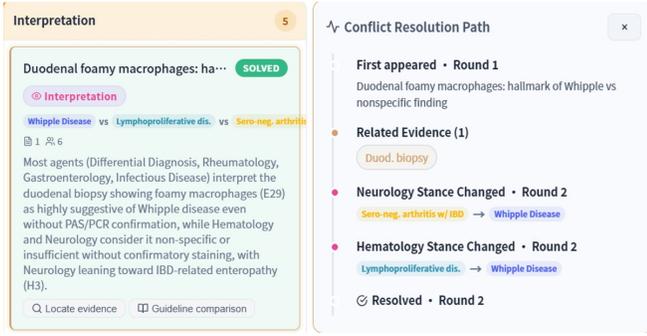

**Figure 4: A resolved conflict card.** The card shows the original disagreement between two hypotheses, the agents involved, a summary of how the conflict was interpreted, and action buttons for locating the disputed evidence or comparing guideline usage.

For deeper inspection, a resolution path records each conflict's lifecycle (Figure 4), including when it first appeared, which agents changed their opinions in response, and when it was resolved, helping clinicians understand how disagreements were eventually settled (**DC5**).

## 5 Evaluation

In this section, we will introduce how we conduct a user study with 12 clinicians to evaluate the effectiveness of our system and the evaluation results.

### 5.1 Setup

To evaluate whether *MDTRoom* effectively supports clinician–agent collaboration in rare disease diagnosis, we conducted a controlled user study with 12 participants (P1–P12; 8 female, 4 male) who had relevant clinical backgrounds (Table 2). Participants ranged in age from 20 to 46 years ($M = 26.83$, $SD = 7.76$) and came from diverse medical specialties. They were recruited through advertisements on social media platforms. Furthermore, only participants who were both familiar with the two cases and able to carry out the diagnostic tasks were recruited. The two cases, designed by an associate chief physician with over 20 years of clinical experience, were selected to ensure comparable difficulty. The study aimed to address the following research questions:

**RQ1:** Does *MDTRoom* improve clinicians' understanding of multi-agent diagnostic discussions?

**RQ2:** Does *MDTRoom* enhance clinicians' sense of control over the multi-agent reasoning process?

**RQ3:** How does *MDTRoom* affect clinicians' engagement and willingness to collaborate with multi-agent diagnostic systems?

**RQ4:** Does *MDTRoom* reduce the cognitive load associated with reviewing and guiding multi-agent discussions?

**Table 2: Participant demographics in the user study.**

| ID  | Gender | Age | Specialty                    | Experience |
|-----|--------|-----|------------------------------|------------|
| P1  | M      | 20  | Radiology                    | 1 year     |
| P2  | F      | 22  | Integrative Medicine         | 3 years    |
| P3  | F      | 25  | Respiratory Medicine         | 6 years    |
| P4  | F      | 22  | Integrative Medicine         | 3 years    |
| P5  | F      | 24  | Ophthalmology                | 5 years    |
| P6  | F      | 21  | Traditional Chinese Medicine | 3 years    |
| P7  | F      | 25  | Respiratory Medicine         | 6 years    |
| P8  | F      | 26  | Translational Medicine       | 7 years    |
| P9  | M      | 46  | Urology                      | 15 years   |
| P10 | M      | 39  | Infectious Diseases          | 15 years   |
| P11 | F      | 25  | Medical imaging              | 6 years    |
| P12 | M      | 27  | Oral Medicine                | 8 years    |

*5.1.1 Baseline.* As a comparison condition, we implemented a reduced interface, termed the Baseline, that retained the same model and agent setup as *MDTRoom* but removed the core components. Specifically, patient information and agent responses were displayed in one linear transcript, without dedicated views for evidence, round-by-round discussion, or conflicts. Participants could respond only through text, rather than directing selected feedback to specific AI specialists.

*5.1.2 Procedure.* We used a counterbalanced within-subjects design. Each participant completed two diagnostic tasks. One task used *MDTRoom*, and the other used the Baseline. The order of conditions and the assignment of cases were counterbalanced across participants to reduce learning and order effects. Each session began with a brief introduction to the multi-agent diagnostic setting and the idea of MDT-style consultation. Participants then received a short orientation to the interface of the condition they used first, followed by a brief familiarization period. They next completed two task sessions in sequence. In each task session, participants reviewed the assigned patient case, observed and interacted with an agent discussion generated live during the session under the assigned condition, and were asked to arrive at a working diagnosis. Participants were encouraged to think aloud throughout. After each task session, they completed a post-task questionnaire and then took part in a semi-structured interview about their strategies, experiences, and preferences. Interviews were audio-recorded with participants' consent, and each participant received approximately US$15 as compensation. This study was approved by the institutional ethics committee.

### 5.2 Quantitative Results

All 12 participants completed both conditions. When asked for an overall preference at the end of the study, all 12 chose *MDTRoom* over the Baseline. Below we report the quantitative comparisons organized by research question (Figure 5). Specifically, we used the Wilcoxon signed-rank test [47] for comparisons between conditions. We report effect sizes as $r = Z/\sqrt{N}$ [13].

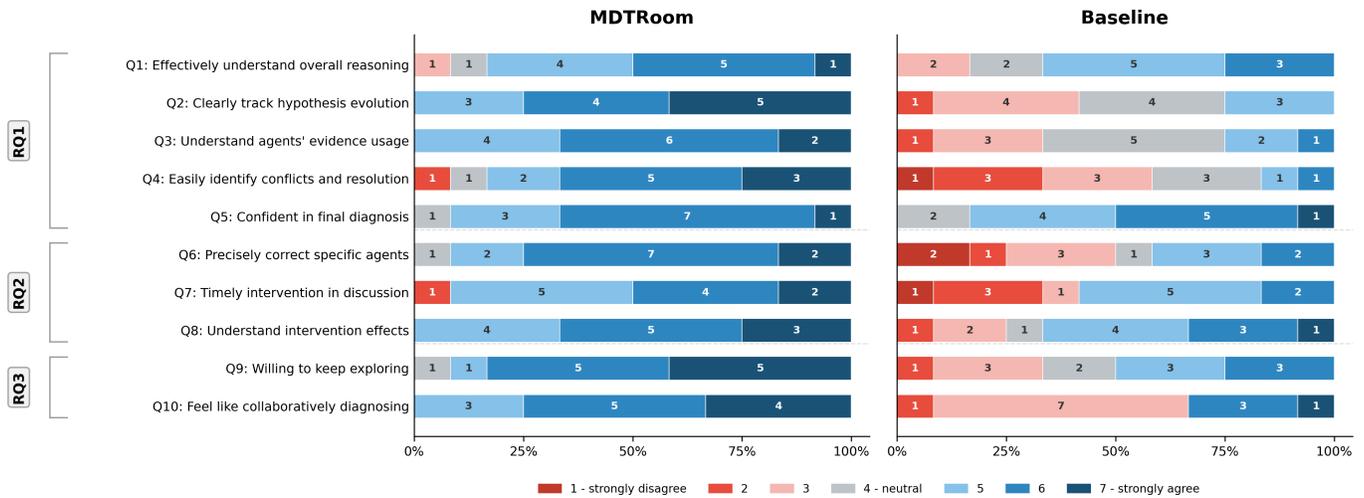

Figure 5: User ratings on *MDTRoom* and the Baseline with a 7-point Likert scale. Q1–Q5 address understanding of multi-agent discussion (RQ1), Q6–Q8 address sense of control (RQ2), and Q9–Q10 address engagement (RQ3).

*5.2.1 Agent Discussion Understanding (RQ1).* We assessed participants' understanding of multi-agent reasoning through both per-condition and direct-comparison items (Q1–Q5 in Fig. 5). On the per-condition item asking whether participants felt they understood the overall reasoning process (Q1), ratings trended higher under *MDTRoom* ($M = 5.33, SD = 1.07$) than the Baseline ($M = 4.75, SD = 1.06; p = .250$). On tracking how agents' opinions evolved across rounds (Q2), *MDTRoom* was rated substantially higher ($M = 6.17, SD = 0.83$) than the Baseline ($M = 3.75, SD = 0.97; p < .001, r = 0.95$). A similarly large effect appeared for understanding how agents used clinical evidence and guidelines in their reasoning (Q3; $M = 5.83$ vs. $3.92; p = .003, r = 0.86$). Participants also found it markedly easier to identify why inter-agent conflicts arose and which remained unresolved (Q4; $M = 5.58$ vs. $3.25; p = .004, r = 0.83$). All three effect sizes exceeded $r = 0.80$, indicating that the structured workspace meaningfully changed how well participants could inspect the reasoning process. On diagnostic confidence (Q5), participants' ratings were slightly higher under *MDTRoom* ($M = 5.67, SD = 0.78$) than the Baseline ($M = 5.42, SD = 0.90; p = .531$).

*5.2.2 Sense of Control (RQ2).* Participants reported substantially greater ability to guide individual agents' reasoning under *MDTRoom* (Q6–Q8 in Figure 5). On targeted correction (Q6), the per-condition item asking whether participants could correct a specific agent without restarting the entire discussion, *MDTRoom* scored $M = 5.83$ ($SD = 0.83$) versus the Baseline's $M = 3.67$ ($SD = 1.78$; $p = .004, r = 0.83$). The final comparative item asking which condition allowed more timely intervention during the discussion (Q7) also favored *MDTRoom* ($M = 5.42$ vs. $3.92; p = .031, r = 0.62$). For understanding the downstream effect of an intervention (Q8), *MDTRoom* again led to higher ratings ($M = 5.92, SD = 0.79$) compared to the Baseline ($M = 4.75, SD = 1.48; p = .039, r = 0.60$).

*5.2.3 Engagement (RQ3).* Across both engagement items, *MDTRoom* elicited significantly higher ratings than the Baseline (Q9–Q10 in Figure 5). Participants reported greater willingness to continue exploring the agent discussion beyond what the task required (Q9; $M = 6.17$ vs. $4.33; p = .008, r = 0.77$). They also felt more like active collaborators in a diagnostic process, rather than passive reviewers of AI output (Q10; $M = 6.08$ vs. $4.00; p = .004, r = 0.83$).

*5.2.4 Cognitive Load (RQ4).* We assessed perceived workload using a 7-point NASA-TLX [17], with lower scores indicating less load. *MDTRoom* significantly reduced mental demand ($M = 3.83, SD = 1.40$ vs. Baseline $M = 5.42, SD = 1.56; p = .015, r = 0.70$) and frustration ($M = 1.58, SD = 0.79$ vs. $M = 2.83, SD = 1.34; p = .016, r = 0.70$). The remaining four NASA-TLX dimensions did not differ significantly between conditions: physical demand ($M = 2.42$ vs. $3.50; p = .071$), temporal demand ($M = 3.00$ vs. $4.33; p = .098$), effort ($M = 3.67$ vs. $4.33; p = .201$), and self-rated performance ($M = 4.67$ vs. $4.67; p = .938$). Nevertheless, overall NASA-TLX scores were lower for *MDTRoom*, suggesting that participants experienced a generally reduced workload compared to the baseline system.

*5.2.5 Feature Usability.* We collected usability ratings (7-point scale) for nine individual components of *MDTRoom* (Figure 6). The highest-rated components were the Evidence Panel (V1; $M = 6.25$), the Clinical Correction tool (V8; $M = 6.25$), and the Agent Detail Modal (V9; $M = 6.17$), all of which serve as direct bridges between patient data and agent reasoning. The Conflict View (V5; $M = 6.00$), the Consensus Summary (V6; $M = 6.00$), and the Agent Discussion View (V3; $M = 5.92$) also received consistently positive evaluations. The Guideline Trace Matrix (V2; $M = 5.25$) and the Round Comparison View (V7; $M = 5.25$) received moderate scores, suggesting room for improvement in how structured guideline references and temporal comparisons are presented. The Hypothesis Evolution View (V4; $M$ = 4.50) received the most mixed ratings, with individual scores spanning from 1 to 7. Several participants noted during the interview that the Sankey-style flow diagram was informative for grasping overall convergence but could become visually cluttered when many competing hypotheses were active.

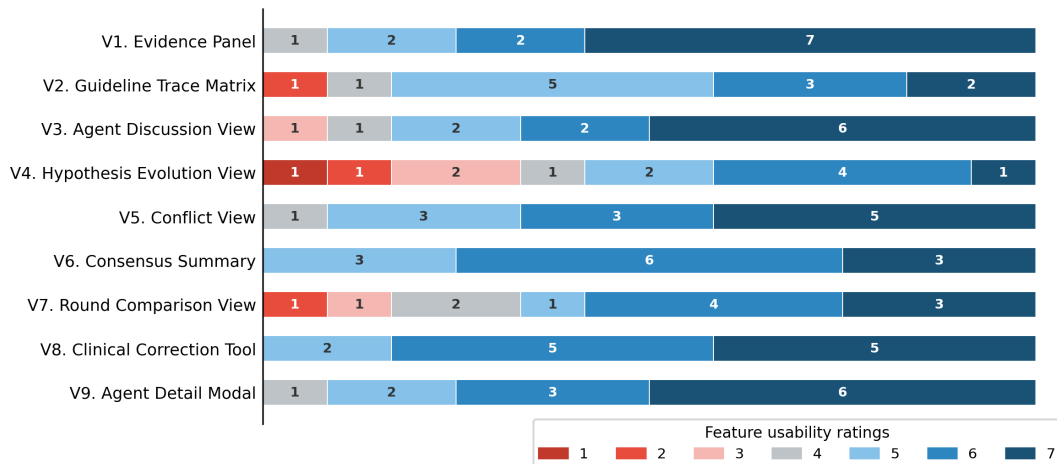

Figure 6: Usability ratings for individual components of *MDTRoom* (7-point scale).

### 5.3 Qualitative Insights

For qualitative analysis, interview recordings were transcribed and analyzed using thematic analysis [4]. Two researchers independently reviewed the transcripts, generated initial codes, and then compared and reconciled their annotations through discussion and by referring back to the recordings. Related codes were iteratively grouped into higher-level themes organized around the four research questions. We report these themes alongside the quantitative results and include representative participant quotations as supporting evidence.

**Surfaced conflicts help clinicians rapidly locate atypical indicators (6/12).** Six participants reported that the system's conflict annotations drew their attention to ambiguous or atypical clinical features that they might not have independently recognized as diagnostically significant. Rather than having to assess every data item exhaustively, clinicians could focus on the subset where specialist agents disagreed, allowing them to better apply their domain knowledge in interpreting the indicators. P8 noted, *"The disagreement points it flagged, such as bone marrow biopsy results and ANA indicators, are precisely the core reasons this case is hard to diagnose. By highlighting these areas, I can focus my attention on the most critical features."* P10 similarly observed that *MDTRoom* narrowed the diagnostic scope and provided a useful reference for exclusionary reasoning.

**Evidence provenance annotations change clinicians' review workflow (9/12).** Nine participants described a fundamental shift in how they approached the agent discussion when evidence provenance was visible. Under the Baseline, they reported that they would read through agent reasoning first and then attempt to trace claims back to the patient data, a process they found laborious and unreliable because raw data was buried within unstructured text. With *MDTRoom*, the workflow was inverted: clinicians began from the highlighted data and evidence items, especially those marked as conflicted, and used their own domain knowledge to evaluate the data before consulting agent interpretations. For instance, P2 explained, *"The citation highlighting on the left lets me locate directly, and clicking into the reasoning shows exactly which item they were reasoning about."* This data-first strategy allowed clinicians to better leverage their clinical expertise more directly.

**The card-based layout enables hypothesis-driven selective inspection (7/12).** Seven participants described entering the discussion workspace with a preliminary diagnosis in mind and then selectively reviewing agents whose conclusions aligned with or diverged from their own hypothesis, rather than reading all agents sequentially. P1 noted, *"By clicking into each agent's dialogue card, I check whether its reasoning chain aligns with my own reasoning."* P5 described a similar strategy of using the system to narrow the diagnostic scope and then verifying the final judgment independently, adding that *"the final decision must still be made by the doctor."* However, under the Baseline, where agent outputs were interleaved in a single transcript, such selective reading was not feasible.

**Tracing consensus retrospectively guides diagnostic focus (4/12).** Four participants reported that they first checked the final consensus or the last round's conclusions, then traced backward to identify which agents had changed their stance and at which round. P11 described this pattern in terms that mirrored real MDT practice: *"Just like a real MDT: first round everyone reviews from their department; second round they exchange views; third round they synthesize a conclusion."* P6 was particularly interested in opinion shifts, noting *"If a doctor initially holds one view then suddenly changes, I really want to know why and what the basis was."* This reverse navigation pattern suggests that stance transitions were perceived as high-signal diagnostic events, and clinicians treated them as clues to the pivotal evidence that resolved (or failed to resolve) the debate.

**Externalized conflicts signaled when clinician expertise was most needed (6/12).** Six participants indicated that the conflict view served not only as a comprehension aid but also as an intervention trigger: it told them precisely when and where their domain knowledge was most needed, rather than requiring them to monitor all agent reasoning continuously. P6 noted, *"If disagreement remains, it actually makes the clinician more prepared. They think about the condition more carefully and rigorously."* P4 described providing her own interpretation for a conflicted finding, which then strengthened the agent's subsequent reasoning and her own confidence in the collaborative diagnosis process.

**Structured layout reduces perceived cognitive load (9/12).** Nine participants explicitly compared the cognitive demands of the two conditions. P10 summarized the contrast starkly: *"The gap is huge. This interface feels like reading a report, a dense mass of text. The first one lets you target specific evidence and doctors, more like discussing a case with an actual colleague."* P9 echoed this sentiment, describing the Baseline as *"scattered text"* where finding cited evidence and department conclusions required *"digging through text piles, which is tiring."* Several participants highlighted specific structural features that contributed to the reduction: P5 noted that the system distilled key abnormalities into card form, greatly reducing reading workload, while P6 valued the ability to selectively click into agent cards rather than parsing dense paragraphs. P12 connected the structured layout to clinical workflow requirements: *"In clinical practice, we need tools that let us see results and key points at a glance, not long-winded reasoning."* These accounts are consistent with the significant reductions in mental demand and frustration observed in the NASA-TLX ratings.

## 6 Discussion

In this section, we discuss our design implications. Furthermore, we elaborate on the limitations in our system with future work.

### 6.1 From Reasoning Engines to Shared Workspaces

Traditional LLM-assisted diagnostic systems often treat multi-agent reasoning as a backend engine for producing final answers and exposing clinicians only to linear, text-heavy reasoning logs. While such logs may preserve completeness, they offer limited support for timely inspection or meaningful intervention in real clinical settings. This design implicitly positions clinicians as downstream consumers of AI outputs, hindering them from effectively guiding and reconstructing agent reasoning trajectories. As multi-agent systems grow more complex, this separation between internal reasoning and human oversight increasingly leads to cognitive overload and reduced opportunities for effective collaboration. Our system reframes multi-agent systems as a shared reasoning workspace, where agent deliberation is externalized into structured, perceptible, and persistent artifacts. By transforming non-linear agent debates into coordinated representations (e.g., evidence groupings, hypothesis evolution, and inter-agent disagreements), the system allows clinicians to fluidly shift between high-level decision-making and detailed auditing. This paradigm enables more timely clinician involvement by grounding human–AI collaboration in shared reference points, supporting joint inspection and alignment.

### 6.2 Reorienting Agent Outputs to Support Clinicians' Use of Domain Knowledge

In many existing multi-agent diagnostic systems, clinicians are required to actively follow individual agent reasoning trajectories in order to make sense of system outputs. This places the burden on clinicians to adapt to agent-centric logic—tracing what a particular agent considered, why certain evidence was emphasized, and how intermediate conclusions were formed. As a result, effective use of the system often depends on clinicians aligning themselves with how agents reason, rather than engaging in diagnosis through their own established clinical workflows. *MDTRoom* shifts this relationship by extracting and externalizing key reasoning information, such as patient data usage and inter-agent conflicts, into standalone and structured artifacts. Rather than inspecting agent-specific reasoning chains, clinicians can engage with these artifacts to directly leverage their own domain knowledge in clinical reasoning. For example, surfaced conflicts help clinicians quickly detect non-typical clinical features that warrant closer attention. Clinicians can first assess these features using their own domain knowledge and, when necessary, subsequently consult agent reasoning to understand how different interpretations were formed. In this way, agent outputs are reoriented from explanations of agent behavior to resources that support clinicians' own sensemaking processes.

### 6.3 Surfacing Data, Evidence, and Hypotheses in Agent Discussions for Diagnosis

Clinical decision-making often unfolds under severe time constraints and cognitive burden. The central challenge is not the availability of information, but determining which information from agent discussions merits attention at a given moment. Our system emphasizes clinically meaningful abstractions grounded in patient data, supporting evidence, and evolving hypotheses, which are the core elements clinicians routinely consider in evidence-based differential diagnosis. The three coordinated views make explicit how raw data are referenced, how evidence is invoked and compared, and how diagnostic hypotheses emerge and evolve during agent deliberation, as well as the relationships among these elements. Looking forward, future work could more deeply integrate how clinicians iteratively adjust their mental models around data, evidence, and hypotheses during diagnosis. By aligning the organization and integration of these core elements with real clinical reasoning processes, multi-agent systems could improve their performance and better support alignment with clinicians, while also providing a foundation for more context-aware interactions.

### 6.4 Limitations and Future Work

This work has several limitations that suggest directions for future research. First, our design primarily targets rare and complex disease settings, where diagnostic uncertainty is high, and clinicians are more willing to engage with agent deliberation to explore alternative hypotheses. In more routine clinical scenarios, clinicians may prioritize efficiency and prefer concise recommendations over inspecting agent discussions, which may limit the applicability of *MDTRoom*'s interaction paradigm. Future work could investigate adaptive interfaces that modulate the visibility and granularity of agent reasoning based on case complexity, clinician intent, or time constraints. Second, *MDTRoom* currently focuses on structured EHR and textual data of patients, where clinical features are relatively well-defined and easier to externalize as reasoning structures. Supporting interaction with more complex data types, such as medical imaging or pathology slides, would require more advanced representations to capture how visual evidence is interpreted, compared, and integrated during agent discussion, which remains an important direction for future work.

## 7 Conclusion

To support effective clinician involvement in LLM-driven multidisciplinary diagnostic debates for rare diseases, we presented *MDTRoom*, an interactive system that transforms multi-agent reasoning from linear, opaque transcripts into a structured and inspectable workspace. Grounded in a formative study with clinicians, *MDTRoom* externalizes patient data, evidence usage, hypothesis evolution, and inter-agent conflicts as first-class interactive objects, enabling clinicians to follow distributed reasoning, intervene in a targeted manner, and guide diagnostic deliberation in real time. Rather than treating multi-agent diagnosis as a fully automated process, *MDTRoom* reframes it as a situated, collaborative activity in which human expertise can be injected precisely when and where it is most needed. Our user study demonstrates that this design meaningfully improves clinician–agent collaboration. In the future, we hope to extend this work toward broader clinical settings and other high-stakes domains, exploring how structured, inspectable interaction paradigms can support human–agent collaboration.